\documentstyle[12pt]{article}
\newcommand{\be}{\begin{equation}}
\newcommand{\ee}{\end{equation}}
\newcommand{\bea}{\begin{eqnarray}}
\newcommand{\eea}{\end{eqnarray}}
\newcommand{\ba}{\begin{array}}
\newcommand{\ea}{\end{array}}
\newcommand{\beas}{\begin{eqnarray*}}
\newcommand{\eeas}{\end{eqnarray*}}
\newcommand{\bes}{\begin{equation*}}
\newcommand{\ees}{\end{equation*}}

\def\i2           {\mbox{$\frac{i}{2}$}}

\begin{document}

\title{\bf Shape invariance method for quintom model in the bent brane background}

\author{{M. R. Setare  $^{a}$\thanks{Email: rezakord@ipm.ir}\hspace{1mm}
J. Sadeghi $^{b}$ \thanks{Email: rezakord@ipm.ir}, A. R. Amani
$^{b}$}\thanks{Email: a.amani@umz.ac.ir}\\
{ $^a$ Department of Science, Payame Noor University, Bijar, Iran}\\
{ $^{b}$ Sciences Faculty, Department of Physics, Mazandaran
University,}\\{  P .O .Box 47415-416, Babolsar, Iran}}

\maketitle
\begin {abstract}
In the present paper, we study the braneworld scenarios in the
presence of quintom dark energy coupled by gravity. The first-order
formalism for the bent brane (for both de Sitter and anti-de Sitter
geometry), leads us to discuss the shape invariance method in the
bent brane systems. So, by using the fluctuations of metric and
quintom fields we obtain the Schrodinger equation. Then we factorize
the corresponding Hamiltonian in terms of multiplication of the
first-order differential operators. These first-order operators lead
us to obtain the energy spectrum with the help of shape invariance
method.
\end {abstract}
\newpage
\section{Introduction}
 It is known that all analytically solvable
potentials in quantum mechanics have the property of shape
invariance~\cite{gend}. In fact shape invariance is an integrability
condition, however, one should emphasize that shape invariance is
not the most general integrability condition as not all exactly
solvable potentials seem to be shape invariance
to~\cite{cooper,dabro}. An interesting feature of supersymmetric
quantum mechanics is that for a shape invariant system
\cite{cooper1, infeld} the entire spectrum can be determined
algebraically without ever referring to underlying differential
equations.\\
In the present paper we would like to use this method for an
interesting problem in cosmology. Here we consider the quintom model
of dark energy \cite{quintom}  in the bent brane background. One of
the most important problems of cosmology, is the problem of
so-called dark energy. The type Ia supernova observations suggests
that the universe is dominated by dark energy with negative pressure
which provides the dynamical mechanism of the accelerating expansion
of the universe \cite{{per},{gar},{ries}}. The strength of this
acceleration is presently matter of debate, mainly because it
depends on the theoretical model implied when interpreting the data.
Most of these models are based on dynamics of a scalar
\cite{{rat},{zlat}, {pha},{tac}} or multi-scalar fields
\cite{quintom}. Primary scalar field candidate for dark energy was
quintessence scenario \cite{{rat},{zlat}}, a fluid with the
parameter of the equation of state lying in the range, $-1< w< {-1
\over 3}$. The analysis of the properties of dark energy from recent
observations mildly favor models with $w$ crossing -1 in the near
past. Meanwhile for the phantom model\cite{pha} of dark energy which
has the opposite sign of the kinetic term compared with the
quintessence in the Lagrangian, one always has $w\leq -1$. Neither
the quintessence nor the phantom alone can fulfill the transition
from $w>-1$ to $w<-1$ and vice versa. But one can show
\cite{quintom} that considering the combination of quintessence and
phantom in a joint model, the transition can be fulfilled. This
model, dubbed quintom, can produce a better fit to the data than
more familiar models with $w\geq-1$.\\
This paper is organised as follows. In section 2 we consider the
quintom model of dark energy in the background of a five-dimensional
space-time with warped geometry. Then we consider the fluctuations
of the metric and quintom scalar fields. In section 3 we review the
supersymmetry algebra with the central charge and shape invariance
method. In section 4 we obtain the factorized Hamiltonian for the
bent brane, which leads us to investigate the shape invariance
method with considering the central extended algebra. Finally by
takeing  advantage of shape invariance method we obtain energy
spectrum in our interesting geometry.
\section{The quintom model in the bent brane background }
The quintom model of dark energy \cite{quintom} is of new models
proposed to explain the new astrophysical data, due to transition
from $w>-1$ to $w<-1$, i.e. transition from quintessence dominated
universe to phantom dominated universe. Containing the normal scalar
field $\phi$ and negative kinetic scalar field $\chi$, the action
which describes the quintom model is expressed as the following form
\begin{equation}\label{s1}
S=\int d^{4}{x} dy
\sqrt{|g|}(-\frac{1}{4}R+{\frac{1}{2}\partial_{a}\phi
\partial^{a}\phi} - {\frac{1}{2}\partial_{a}\chi
\partial^{a}\chi}
- V(\phi,\chi)).
\end{equation}
where we have not considered the lagrangian density of matter field,
and we take $4{\pi}{G}=1$. The line element of the five-dimensional
space-time can be written
 \begin{equation}\label{s2}
{ds}_{5}^{2}=g_{ab}{dx}^{a}{dx}^{b}=e^{2A}{ds}_{4}^{2}-{dy}^{2}
\end{equation}
where $a, b = 0, 1, 2, 3, 4$, and $e^{2A}$ is the warp factor.
$dS_{4}^{2}$ represent the four-dimensional metric:
\begin{equation}\label{s3}
{ds}_{4}^{2}={dt}^{2}-e^{2\sqrt{\Lambda}t}({dx}_{1}^{2}+{dx}_{2}^{2}+{dx}_{3}^{2})
\end{equation}
where $\Lambda$  is four dimensional cosmological constant. We note
that constant $\Lambda$ is positive for de Sitter $(dS_{4})$
spacetime, negative for anti - de Sitter $(AdS_{4})$ spacetime and
zero for Minkowski $(M_4)$ spacetime.\\   At first we consider the
interacting case with $\Lambda = 0$, also  the functions $A$,
$\phi$, $\chi$ are
$A(y)$,$\phi(y)$,$\chi(y)$.\\
From the Einstein and Euler–Lagrange equations we obtain,
\begin{eqnarray}\label{s4}
{A''}&=& -\frac{2}{3}({\phi '}^2-{\chi '}^2),\nonumber\\
{A'}^{2} &=&\frac{1}{6}({\phi '}^2-{\chi'}^2)-\frac{1}{3}V(\phi,\chi),\nonumber\\
 V_{\phi}&=& {\phi''}+4{A'}{\phi '} ,\nonumber\\
 V_{\chi}&=&-{\chi''}-4{A'}{\chi '}
\end{eqnarray}
where a prime denotes a derivative with respect to $y$, and
\begin{equation}\label{ss3}
V_{\phi}=\frac{dV}{d\phi},\hspace{1cm} V_{\chi}=\frac{dV}{d\chi}.
\end{equation}
In order to obtain the first-order equation, we use \cite{baz}
\begin{eqnarray}\label{s5}
{A'} &=&-\frac{1}{3}W,\nonumber\\
{\phi '} &=& \frac{1}{2}W_{\phi},\nonumber\\
 {\chi '} &=& -\frac{1}{2}W_{\chi}
\end{eqnarray}
From (\ref{s4}) and (\ref{s5}) the explicit form of the potential is
\begin{equation}\label{s6}
V(\phi,\chi)=\frac{1}{8}({W_{\phi}}^2-{W_{\chi}}^2)-\frac{1}{3}W^2
\end{equation}\\Next we consider the general case with $\Lambda\neq0$ and we obtain

\begin{eqnarray}\label{s7}
{A''}+{\Lambda}e^{-2A} &=& -\frac{2}{3}({\phi '}^2-{\chi '}^2),\nonumber\\
{A'}^{2}-{\Lambda}e^{-2A} &=& \frac{1}{6}({\phi '}^2-{\chi
'}^2)-\frac{1}{3}V(\phi,\chi)
\end{eqnarray}
The cosmological constant leads us to define the function which
corresponds to the scalar fields $\phi$ and $\chi$. It means that
this function is completely coupled and generally responsible for
the cosmological constant. Thus we gain
\begin{eqnarray}\label{s8}
 {A'} &=& {-\frac{1}{3}W-\frac{1}{3}{\Lambda}{\gamma}Z},\nonumber\\
{\phi '} &=& \frac{1}{2}W_{\phi}+\frac{1}{2}{\Lambda}{(\alpha+\gamma)}Z_{\phi},\nonumber\\
 {\chi '} &=& -\frac{1}{2}W_{\chi}-\frac{1}{2}{\Lambda}{(\gamma+\beta)}Z_{\chi},
\end{eqnarray}
where $Z=Z(\phi,\chi)$ is a new and arbitrary function of the scalar
fields and respond  for the presence of the cosmological constant.
$\alpha,$ $\beta$ and $\gamma$ are constants.  The corresponding
potential is,
\begin{eqnarray}\label{s10}
V(\phi,\chi)&=&\frac{1}{8}({W_{\phi}}+{\Lambda}(\alpha+\gamma)Z_{\phi})({W_{\phi}}+{\Lambda}(\gamma-3\alpha)Z_{\phi})-
\frac{1}{3}(W+{\Lambda}\gamma Z)^2
 \nonumber\\ &- &\frac{1}{8}({W_{\chi}}+{\Lambda}(\beta+\gamma)Z_{\chi})({W_{\chi}}+{\Lambda}(\gamma-3\beta)Z_{\chi})
\end{eqnarray}
Here we assume that
\begin{equation}
W_{\phi\chi} = 0
\end{equation}
So, by inserting this potential in the equations of motion one can
obtain the following constraint,
\begin{equation}\label{s11}
\alpha W_{\phi\phi}Z_{\phi}+\alpha
W_{\phi}Z_{\phi\phi}+2{\Lambda}\alpha(\alpha+\gamma)Z_{\phi}
Z_{\phi\phi}-\frac{4}{3}\alpha Z_{\phi}(W+\Lambda\gamma Z)=0
\end{equation}
\begin{equation}\label{s11}
\beta W_{\chi\chi}Z_{\chi}+\beta
W_{\chi}Z_{\chi\chi}+2{\Lambda}\beta(\beta+\gamma)Z_{\chi}Z_{\chi\chi}+\frac{4}{3}\beta
Z_{\chi}(W+\Lambda\gamma Z)=0
\end{equation}
For simplicity, we  consider $Z(\phi, \chi)=W(\phi, \chi).$ This
possibility leads to equations,
\begin{equation}\label{s12}
\frac{3}{2}{d}{W}_{\phi\phi}-W = 0,
\end{equation}
\begin{equation}\label{s13}
 -\frac{3}{2}{d^\prime}{W}_{\chi\chi}+W
= 0
\end{equation}
where $d=\frac{1+\Lambda(\gamma+\alpha)}{1+\Lambda\gamma}$ and
$d^{\prime}=-\frac{(1+\Lambda(\gamma+\beta))}{1+\Lambda\gamma}$. If
we consider the combination of Eqs.(\ref{s12}, \ref{s13}) we obtain:
\begin{equation}\label{phi-sig}
dW_{\phi\phi}-d^{\prime}W_{\chi\chi}=0
\end{equation}
To explain this new result, we take  super-potential as,
\begin{equation}\label{phi-sig1}
W(\phi,\chi )= 3a\sinh\left(b\phi+c\chi\right)
\end{equation}
where $b= \sqrt{\frac{2}{3d}}$ and $c=\sqrt{\frac{2}{3d^{\prime}}}$,
and $a$ is a constant. By substituting super-potential
(\ref{phi-sig1}) into Eqs.(\ref{s10}), (\ref{s8}) one can obtain
following equations respectively for potential, scalar fields, and
$A(y)$
\begin{eqnarray}\label{phi-sig2}
V(\phi,\chi)=-\frac{3}{4}a^2
(1+\Lambda\gamma)\left(2+\Lambda(2\gamma+3\alpha+3\beta)\right)\cosh^2(b\phi+c\chi)+3a^2(1+\Lambda\gamma)^2
\end{eqnarray}

also the $\phi(y)$, $\chi(y)$ and $A(y)$ as follow,
\begin{equation}
\phi(y)=
\pm\sqrt{\frac{3d}{8}}\ln\left[\tan(a(1+\Lambda\gamma)y)\right]
\end{equation}
\begin{equation}
\chi(y)=\pm\sqrt{\frac{3d^{\prime}}{8}}\ln\left[\tan(a(1+\Lambda\gamma)y)\right]
\end{equation}
\begin{equation}
A(y)=
\frac{1}{2}\ln\left[\frac{1}{2}\sin(2a(1+\Lambda\gamma)y)\right]
\end{equation}
Now we are going to discuss shape invariance condition. This
condition help us to investigate the stability of system. So, first
we consider the fluctuations of the metric and scalar fields.\\ The
perturbed metric is,
\begin{equation}
ds^2=e^{2A}(g_{\mu \nu} + \epsilon h_{{\mu \nu}}) dx^\mu dx^\nu -
dy^2
\end{equation}
We use the coordinate $z$ which is defined by following expression,
\begin{equation}
dz=e^{-A(y)}dy
\end{equation}
So, one can obtain the coordinate $z$ as a follow,
\begin{equation}
z=\frac{1}{a(1+\Lambda\gamma)} \ln
\left[\tan(a(1+\Lambda\gamma)y)\right]
\end{equation}
and $y$ is
\begin{equation} y=\frac{1}{a(1+\Lambda\gamma)}
\arctan(e^{a(1+\Lambda\gamma)z})
\end{equation}
Also, $A(z)$ be in the form,
\begin{equation}
A(z) = \frac{1}{2}\ln\left[\frac{1}{2}sech(\eta z)\right]
\end{equation}
where for simply, $a=1$ and $\eta=(1+\Lambda\gamma)$.
\section{Shape Invariance Method}
If the ground state energy is zero, we can factorize the Hamiltonian
as,
\begin{eqnarray}
H_{1}(g) = B^{\dagger}(g)B(g)
\end{eqnarray}
where $g$ is (are) the real parameter(s),  which give(s) us the
potential, and $B(g)$ is  a first order differential operator. The
ground state of $H_{1}$ is annihilated by $B(g)$. The partner
Hamiltonian of $H_{1}$ will be obtain with reversing the order of
$B$ and $B^{\dagger},$
\begin{equation}
H_{2}(g) = B(g)B^{\dagger}(g).
\end{equation}
The spectrum of $H_{1}$ and $H_{2}$ is degenerate. The only
difference is that $H_{1}$ has a zero-energy state and $H_{2}$ in
general does not, so we have,
\begin{equation}
H_{2}B = BH_{1}.
\end{equation}
If we had for $n\ge 0,$
\begin{equation}
H_1{\Psi _n}^{(1)}=E_{n}^{(1)}{\Psi_n}^{(1)}
\end{equation}
this implies that,
\begin{equation}
H_{2}(B{\Psi_{n}}^{(1)}) ={E_{n}}^{(1)}(B{\Psi_{n}}^{(1)}).
\end{equation}
S0, the relation between the eigenvalues and eigenfunctions of the
two hamiltonians $H_{1}$ and $H_{2}$ are,
\begin{eqnarray}
E_{n}^{(2)} &=& E_{n+1}^{(1)} ,  \qquad  E_{0}^{(1)} =0,\nonumber\\
\Psi_{n}^{(2)}&\propto &A\Psi_{n+1}^{(1)},
\end{eqnarray}
where the ground state wavefunction for $H_{1}(orH_{2})$ can be
obtained as,
\begin{eqnarray}
B \Psi_{0}^{(1)}(x) = 0 \Rightarrow \Psi_{0}^{(1)}(x) = N\exp(-\int^xW(y)d(y))\nonumber\\
B^{\dagger}\Psi_{0}^{(2)}(x) = 0 \Rightarrow
\Psi_{0}^{(2)}(x)=N\exp(+\int^xW(y)d(y)).
\end{eqnarray}
We know that, supersymmetry give relationships between Hamiltonian
$H_{1}$ and $H_{2}$, where $H_1$ and $H_2$ are partner of each
other.
\begin{equation}\label{s15}
H=\left(\begin{array}{ccc}
H_{1} & 0 \\
0 & H_{2} \\
\end{array}\right).
\end{equation}
This matrix can be obtained from anticommutator $H=\{Q,Q\dagger\}$,
where $Q$ and $Q\dagger$ are supercharges, given by,
\begin{equation}
Q=\left(\begin{array}{ccc}
0 & 0 \\
B & 0 \\
\end{array}\right),\qquad
Q^{\dagger}=\left(\begin{array}{ccc}
0 & B^{\dagger} \\
0 & 0 \\
\end{array}\right)
\end{equation}
In this algebra we have
\begin{eqnarray}
[H,Q]&=&[H,Q^\dagger]=0\nonumber\\
\{Q,Q\}&=&\{Q^\dagger,Q^\dagger\}=0
\end {eqnarray}
A potential is said to be shape invariant, if its supersymmetry
partner potential has the same spatial dependence as the original
potential, with possibly altered parameters. Suppose that, these
Hamiltonian are linked by the condition,
\begin{equation}
B(g_1)B^{\dagger}(g_1)=B^{\dagger}(g_2)B(g_2)+c(g_2),
\end{equation}
where the real parameters $g_1$ and $g_2$ are related by a mapping
$f:g_1 \longrightarrow g_2$, and $c(g)$ is a c-number that depends
on the parameter(s) of the Hamiltonian, so we have,
\begin{equation}
H_{k}={B^\dagger}(g_{k})B(g_{k})+c(g_{k})+\cdots+c(g_{2}),
\end {equation}
where
\begin{eqnarray}
g_{j+1}&=&f(g_{j})\nonumber\\
B^\dagger(g_{k})H_{k+1}&=&H_{k}B^{\dagger}(g_{k})
\end{eqnarray}
The ground state of each of these sectors satisfies  a first-order equation,namely $$B(g_{k})\Psi_{1}(x;g_k)=0.$$\\
Now, we study supersymmetry with central charge. Supersymmetric
quantum mechanics [10] can be formulated as a one - dimensional
supersymmetric quantum field theory. A bosonic field is, then, a
real - valued function of time, and a fermionic field is a Grassman
- valued function of time. The $d=1$, $N=1$ superalgebra with a
central charge is specified by the following relations,
\begin{eqnarray}
\{Q,Q^\dagger\}&=&H\nonumber\\
\left[H,Q\right]&=&\left[H,Q^\dagger\right]=0\nonumber\\
\{Q,Q\}&=&\{Q^\dagger,Q^\dagger\}=C,
\end{eqnarray}
where $Q$ and $C$ are supercharge and central charge respectively.
The above algebra implies $[Q, C]=[Q^{\dagger}, C]=0$

To realize the algebra (40), we represent the supercharges as
matrices,
\begin{equation}
Q=\left(\begin{array}{ccc}
\lambda & 0 \\
B & -\lambda \\
\end{array}\right)\qquad
Q^{\dagger}=\left(\begin{array}{ccc}
\lambda & B^{\dagger} \\
0 & -\lambda \\
\end{array}\right)
\end{equation}
where $\lambda$ is a real $c$-number. This approach is first to present an implementation of this algebra in a two - sector model, and then to generalize this construction to an arbitrary number of sectors.\\
The corresponding Hamiltonian  for two-sector is,
\begin{equation}
H=\left(\begin{array}{ccc}
B^{\dagger}B+2\lambda^2 & 0 \\
0 & BB^{\dagger}+2\lambda^2 \\
\end{array}\right),\qquad
C=\left(\begin{array}{ccc}
2\lambda^2 & 0 \\
0 & 2\lambda^2 \\
\end{array}\right),\qquad C\ge 0
\end{equation}
\\To construct a model with 4 sectors , one can concentrate 2 two - sector model. It has supercharges
\begin{equation}
Q=\left(\begin{array}{cccc}
-\lambda_1 & 0 & 0 & 0\\
B_1 & \lambda_1 & 0 & 0\\
0 & 0 & -\lambda_3 & 0\\
0 & 0 & B_3 & \lambda_3
\end{array}\right)\qquad
Q^{\dagger}=\left(\begin{array}{cccc}
-\lambda_1 & {B_1}^{\dagger} & 0 & 0\\
0 & \lambda_1 & 0 & 0\\
0 & 0 & -\lambda_3 & {B_3}^{\dagger} \\
0 & 0 & 0 & \lambda_3
\end{array}\right)
\end{equation}
we obtain,
\begin{equation}
H=\left(\begin{array}{cccc}
{B_1}^{\dagger}B_{1}+2{\lambda_1}^2 & 0 & 0 & 0\\
0 & B_{1}{B_1}^{\dagger}+2{\lambda_1}^2 & 0 & 0\\
0 & 0 & {B_3}^{\dagger}B_{3}+2{\lambda_3}^2  & 0\\
0 & 0 & 0 & B_{3}{B_3}^{\dagger}+2{\lambda_3}^2
\end{array}\right)
\end{equation}
and
\begin{equation}
C=\left(\begin{array}{cccc}
2{\lambda_1}^2 & 0 & 0 & 0\\
0 & 2{\lambda_1}^2 & 0 & 0\\
0 & 0 & 2{\lambda_3}^2  & 0\\
0 & 0 & 0 & 2{\lambda_3}^2
\end{array}\right)
\end{equation}
As we see the sectors 1 and 2 are degenerate,with energies bounded
from below by $2\lambda_{1}^{2}$, and sectors 3 and 4 are
degenerate, with energies bounded from below by $2\lambda_{3}^{2}$.
The only exceptions are that sectors 1 and 3 each have states that
saturate their respective energy bounds while the even sectors do
not.\\ This suggest an enhanced algebraic structure. In four sector
case,we define the shift operator S by
\begin{equation}
S\equiv\left(\begin{array}{cccc}
0& 0 & 0 & 0\\
B_{1} & 0& 0 & 0\\
0 &D&  0 & 0\\
0 & 0 & B_{3} & 0
\end{array}\right)
\end{equation}
We can choose $D$ such that shape invariance and
$\left[H,S\right]=0$ satisfy to corresponding case. For this,  we
suppose a unitary transformation which is  represented by an
operator $\Omega$ such that $B_{3}={\Omega}^{\dagger}B_{1}\Omega$.
And also, we use a unitary operator $U$ such that ${U}^{2}=\Omega$,
and the conserved shift operator takes the form,
\begin{equation}
S\equiv\left(\begin{array}{cccc}
0 & 0 & 0 & 0\\
B_{1} & 0 & 0 & 0\\
0 & U^{\dagger}B_{1}U &  0 & 0\\
0 & 0 & {U^{\dagger}}^{2} B_{1}{U}^{2} & 0
\end{array}\right),
\end{equation}
From conservation of S, one can obtain the shape invariance
relation,
\begin{eqnarray}
B_{1}{B_{1}}^{\dagger}-{U}^{\dagger}{B_{1}}^{\dagger}B_{1}U=k,\nonumber\\
2{\lambda_{3}}^{2}=2{\lambda_{1}}^{2}+k+{U}^{\dagger}kU.
\end{eqnarray}
Therefore, the Hamiltonian of the four sector model is related to S
in an especially simple way. In particular
\begin{equation}\label{eq49}
H={S}^{\dagger}S+F,
\end{equation}
where
\begin{equation}\label{eq50}
F=\left(\begin{array}{cccc}
2{\lambda_1}^2 & 0 & 0 & 0\\
0 & 2{\lambda_1}^2+k & 0 & 0\\
0 & 0 & 2{\lambda_1}^2+k+{U}^{\dagger}kU  & 0\\
0 & 0 & 0 & H_{4}
\end{array}\right).
\end{equation}
In the first three sectors, the energies are constrained by a
Bogmol'nyi bound, $H_{k}\ge (F)_{kk}$,
because each of the first sector has to be degenerate with the  Bogmol'nyi  - saturating ground state of one of the first three sectors. The constants in $F$ represent not only the Bogmol'nyi bounds of the various sectors, but also the first three energy eigenvalues of the original Hamiltonian.\\
Next section, we take advantage from above information, and  also apply to bent brane with quintom dark energy.\\

\section{The Stability of System  with Shape Invariance Method}
In order to discuss the shape invariance condition we have to make
the corresponding Schr\"odinger equation in terms of coordinate $z$

\begin{equation}\label{s17}
-\frac{d^{2}{\psi}(z)}{{dz}^2}+V(z){\psi}(z)=n^{2}{\psi}(z)\\
\end{equation}
where
\begin{equation}
V(z) = -\frac{9}{4}\Lambda+\frac{3}{2}
A^{\prime\prime}(z)+\frac{9}{4}A^{\prime 2} (z)
\end{equation}
Now we can factorize the corresponding schrodinger equation in the
form
\begin{equation}
\left[\frac{d}{dz}+\frac{3}{2}A^{\prime}(z)\right]\left[-\frac{d}{dz}+\frac{3}{2}A^{\prime}
(z)\right]\psi(z)=(n^2+\frac{9}{4}\Lambda)\psi(z)
\end{equation}
The corresponding potential in term of $z$ is ,
\begin{equation}
V(z) = s^{2} \eta^{2}-s(s+1) \eta^{2} sech^{2}(\eta z)
\end{equation}
where $s=\frac{3}{4}.$\\
By using the new variable  $x=\lambda z$ the  Schr\"odinger equation
 can be written as,
\begin{equation}
-\frac{d^{2}{\psi}(x)}{{dx}^2}+\left[s^2-s(s+1)
sech^{2}(x)\right]{\psi}(x)=(\frac{n^{2}}{\eta^2}+\frac{9\Lambda}{4\eta^2}){\psi}(x)
\end{equation}
and we have
\begin{equation}
H = -\frac{d^2}{dx^2}-s(s+1)\sec h^2(x)+s^{2}
\end{equation}
Now we are going to factorize $H$ in terms of lowering and raising
operators, respectively,
\begin{eqnarray}
B&=&-\frac{d}{dx}-s\tanh(x)\nonumber\\
{B}^{\dagger}&=&\frac{d}{dx}-s\tanh(x),
\end{eqnarray}
and one can obtain the paired Hamiltonians
\begin{eqnarray}
H_{1}&=&{B}^{\dagger}B=-\frac{d^2}{dx^2}-s(s+1)\sec h^2(x)+s^{2}\nonumber\\
H_{2}&=&B{B}^{\dagger}=-\frac{d^2}{dx^2}-s(s-1)\sec h^2(x)+s^{2},
\end{eqnarray}
where
\begin{equation}
H_{2}(s)=H_{1}(s-1)+c(s).
\end{equation}
This relation shows us there is a shape invariance condition with $c(s)=2s-1.$\\
In the case of a central charge, we choose  unitary operator $U$ as
follows
\begin{equation}
U=\exp(\frac{\partial}{\partial s}),\qquad
U^{\dagger}=\exp(-\frac{\partial}{\partial s}),
\end{equation}
where
$$U^{\dagger} f(s)U\longrightarrow f(s-1)$$
From Eqs. (46) and (47) we have,
\begin{equation}
{S}^{\dagger}S=\left(\begin{array}{cccc}
 {B_{1}}^{\dagger}B_{1}& 0 & 0 & 0\\
0 & {U}^{\dagger} {B_{1}}^{\dagger}B_{1}U& 0 & 0\\
0 & 0 &{\Omega}^{\dagger}{B_{1}}^{\dagger}B_{1}\Omega & 0\\
0 & 0 & 0 & H_{4}
\end{array}\right),
\end{equation}
with
\begin{eqnarray}
{B_{1}}^{\dagger}B_{1}&=&-\frac{d^2}{dx^2}-s(s+1)\sec h^2(x)+(s-1)^{2}\nonumber\\
{U}^{\dagger}{B_{1}}^{\dagger}B_{1}U&=&-\frac{d^2}{dx^2}-(g-1)(g)\sec
h^2(x)+(s-2)^{2}\nonumber\\
{\Omega}^{\dagger}{B_{1}}^{\dagger}B_{1}\Omega&=&-\frac{d^2}{dx^2}-(s-2)(s-1)\sec
h^2(x)+(s-3)^{2}.
\end{eqnarray}
Also, by using Eqs.(\ref{eq49}),(\ref{eq50}), one can obtain $F$ as
follows:
\begin{equation}
F=\left(\begin{array}{cccc}
-\frac{9\Lambda}{4\eta^2} & 0 & 0 & 0\\
0 & -\frac{9\Lambda}{4\eta^2}+2s-1 & 0 & 0\\
0 & 0 & -\frac{9\Lambda}{4q^2}+4s-4 & 0\\
0 & 0 & 0 & H_{4}
\end{array}\right)
\end{equation}
Therefore, the energy spectrum of $H_{1}$ is
\begin{eqnarray}
E_{0}^{(1)}&=&(-\frac{9{\Lambda}}{4})\nonumber\\
E_{1}^{(1)}&=&(-\frac{9{\Lambda}}{4}+\frac{1}{2}\eta^2)\nonumber\\
E_{2}^{(1)}&=&(-\frac{9{\Lambda}}{4}-\eta^2)\nonumber\\
E_{3}^{(1)}&=&H_{4}
\end{eqnarray}
So, here we can discuss three cases as $\Lambda=0$, $\Lambda <0$ and
$\Lambda >0$ which are corresponding to flat, AdS and dS space
respectively. The energy spectrum of $H_{1}$ for flat, AdS and dS
spaces are as following respectively. In the another term for
$\Lambda=0$ we have
\begin{eqnarray}\label{fl}
E_{0}^{(1)}&=&0\nonumber\\
E_{1}^{(1)}&=&\frac{1}{2}\eta^2>0\nonumber\\
E_{2}^{(1)}&=&-\eta^2<0\nonumber\\
E_{3}^{(1)}&=&H_{4}
\end{eqnarray}
for $\Lambda <0$ we have
\begin{eqnarray}\label{an}
E_{0}^{(1)}&=&(-\frac{9{\Lambda}}{4})>0\nonumber\\
E_{1}^{(1)}&=&(-\frac{9{\Lambda}}{4}+\frac{1}{2}\eta^2)>0\nonumber\\
E_{2}^{(1)}&=&(-\frac{9{\Lambda}}{4}-(1+\Lambda \gamma)^2)\leq 0 or \geq0\nonumber\\
E_{3}^{(1)}&=&H_{4}>0
\end{eqnarray}
and finally for $\Lambda >0$ we have
\begin{eqnarray}\label{des}
E_{0}^{(1)}&=&(-\frac{9{\Lambda}}{4})<0\nonumber\\
E_{1}^{(1)}&=&(-\frac{9{\Lambda}}{4}+\frac{1}{2}(1+\Lambda \gamma)^2)\leq 0 or \geq0\nonumber\\
E_{2}^{(1)}&=&(-\frac{9{\Lambda}}{4}-\eta^2)<0\nonumber\\
E_{3}^{(1)}&=&H_{4}>0
\end{eqnarray}
We note that in the all above cases we have some tachyonic states
with negative energy. From (\ref{an}) one can see, if
$|\frac{9{\Lambda}}{4}|> (1+\Lambda \gamma)^2$ then for the $AdS_4$
case, all states have positive eigenvalues. In this case the
transition from $AdS_4$ to $M_4$ and $dS_4$ geometry is not stable.
\section{Conclusion}
In the present paper we have described the algebra which gives a
natural framework for understanding the origins of shape invariance
in our interesting problem. The study of shape invariance solutions
can be done by the factorization method. Our aim was to solve and
discuss the stability of a bent brane in the presence of quintom
dark energy and in different geometries with a non-zero cosmological
constant. We have done the perturbation to the metric and fields and
achieved the corresponding Schrodinger equation which was the
second-order equation. Then we have factorized the equation to the
first-order equations which are raising and lowering operators and
have  generated the algebra. From  first order equations we easily
discussed the energy spectrum and also the stability of the system
in the transition to different geometries.

\end{document}